\documentclass{iaus}
\usepackage{graphicx}

\title[Masses of the astrometric SB2 
$\zeta$\,Ori A] 
{Masses of the astrometric SB2 
$\zeta$\,Ori A\footnote{Based on observations
    under ESO programs 076.C-0431, 080.A-9021, and 083.D-0589.}}

\author[Th.\ Rivinius, C.~A.\ Hummel, O.~Stahl]   
{Th.\ Rivinius$^{1}$, C.~A.~Hummel$^2$, O.~Stahl$^3$ 
}

\affiliation{$^1$ESO Chile; 
$^2$ESO, Germany;
$^3$ZAH/LSW Heidelberg, Germany
}

\pubyear{2011}
\volume{272}  
\pagerange{1--2}
\jname{Active OB stars: structure, evolution, mass loss and critical limits}
\editors{C.\ Neiner, G.\ Wade, G.\ Meynet \& G.\ Peters, eds.}
\begin{document}
\maketitle

\begin{abstract}
We report the first dynamic mass for an O-type supergiant, the
interferometrically resolved SB2 system $\zeta$\,Ori A (O9.5Ib+B0/1). The
separation of the system excludes any previous mass-transfer, ensuring that the
derived masses can be compared to single star evolutionary tracks.

\end{abstract}

\firstsection 
\section{Introduction}
Stellar {masses} in the {upper HRD} are notoriously hard to constrain.  Very
few masses are known independently from { stellar evolution} or {wind} models,
mostly using eclipsing SB2 systems.  For stars that have already {evolved}
away from the main sequence, this is a {problematic} technique, though: These
are usually quite narrow short period systems, which means that the
possibility of {mass transfer} via {overflow} having altered the evolutionary
paths is high.  {Interferometry} is in principle able to {overcome this
  problem} in cases where the orbit can be measured for an SB2.

\cite[Hummel et al.~(2000)]{hum00} found the O9.5\,Ib primary of $\zeta$\,Ori
A to be a multiple star. In addition to the well known {B-component} of
spectral type B0 at V=3.77mag, several {arcseconds away}, they found another
companion (Ab) 40\,mas away, about 2\,mag fainter than the primary (Aa),
i.e.\ the magnitude of Ab is similar to that of component B.

\section{Observations}
Additional {Interferometric} observations to complete the orbit coverage
(Fig. 1, right).  were obtained at the NPOI in Flagstaff,
Arizona. {Spectroscopic} observations were taken with the echelle instruments
HEROS and FEROS.

In the spectra all He\,{\sc i} lines as well as He\,{\sc ii} 4686 have a
relatively {narrow core} with varying RV in one sense, while the line {wings}
are shifted in {anti-phase} wrt.\ the cores. This is the signature of an SB2
binary. For some lines (almost) exclusive {formation in the O9.5 component}
can be assumed. The best candidates are the He\,{\sc ii} lines, typically not
seen in B-type stars, except He\,{\sc ii} 4686. The radial velocities of this
line were measured with Gaussian fits to the line center (Fig.~1, left). There
are as well very weak and rather {narrow lines} that are not expected in the
O9\,Ib star.  These narrow lines are {RV variable} in the same sense as the
cores of stronger lines, i.e.~they belong to the companion and are indicative
for an early type B star.

In addition to Gaussian fits, the RV curve was also measured by {spectral
  disentangling} with {VO-KOREL}, a virtual observatory tool based on the
KOREL code by \cite[Hadrava (1995)]{had95}.

\begin{figure}
\centering
\parbox{\textwidth}{%
\parbox{0.5\textwidth}{\includegraphics[viewport=55 39 327 307,angle=0,width=0.5\textwidth,clip]{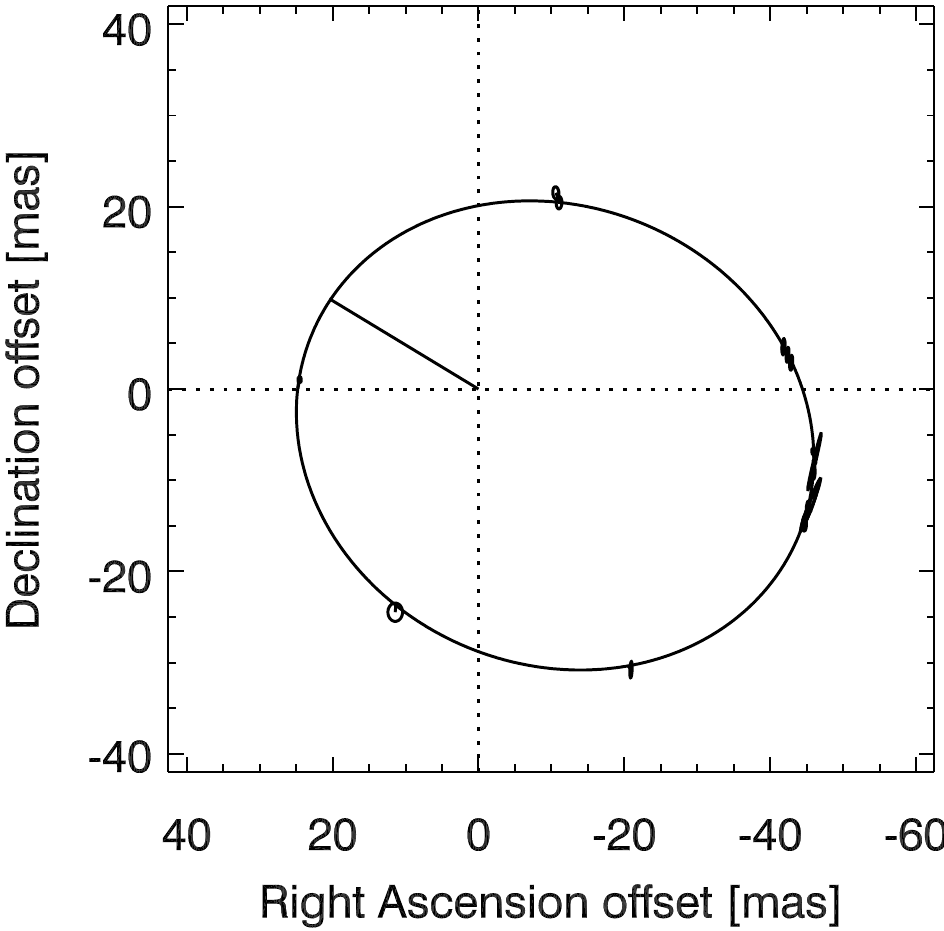}}%
\parbox{0.5\textwidth}{\includegraphics[viewport=55 39 327 307,angle=0,width=0.5\textwidth,clip]{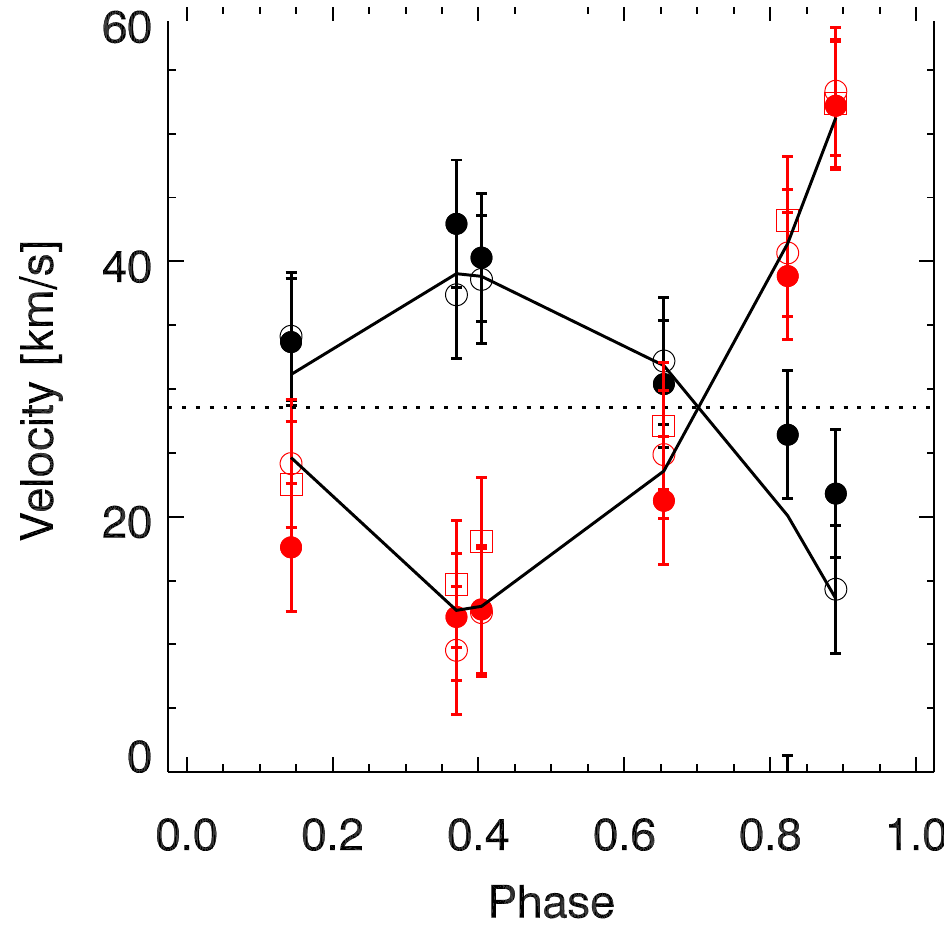}}%
}
\caption[]{
\centering
Left panel: The interferometric orbit. The periastron is marked by the line
from the origin to the orbit. The secondary progresses clockwise.
\newline Right panel: he measured RVs of both components. The filled disks are
Gaussian centers (He\,{\sc ii} and O\,{\sc ii}. The other RVs are from
spectral disentangling of O\,{\sc ii}\,4940 (rectangles) and He\,{\sc i}\,6678
(circles)
}
\label{FNCMa}
\end{figure}

\section{Spectro-interferometric parameters}
The orbital parameters were derived from a {simultaneous fit} to the {\bf
  interferometric} and {spectroscopic} data: \medskip

\noindent
\begin{center}
\begin{tabular}{lr|lr}
$P_{\rm orb.}$ & $2677.5\pm6.9$\,d \hspace*{5mm}& \hspace{5mm}          inclination & $137.9\pm1.1^\circ$\\   
Primary periastron date & JD=$2\,452\,747.9\pm5.2$ \hspace*{5mm}&\hspace{5mm}    semi-major axis & $36.1\pm0.3$\,mas\\ 
Eccentricity & $0.34\pm0.01$ \hspace*{5mm}& \hspace{5mm}                  $q$ & 0.66 \\                         
Periastron long. & $208.4\pm1.1^\circ$ \hspace*{5mm}&\hspace{5mm}      $\gamma$ & 28.6\,km/s \\              
Ascending Node & $86\pm0.8^\circ$ \hspace*{5mm} \\
$M_{\rm Aa}$ & {24.8$\pm$5.6\,M$_\odot$} \hspace*{5mm}&\hspace{5mm} $M_{\rm
  Ab}$ & {16.4$\pm$4.9\,M$_\odot$}\\

\end{tabular}
\end{center}
\vspace*{5mm}

%
This is the {first directly measured (preliminary) mass for an O-type
  supergiant unaffected by mass transfer}, as well being the first application
of the interferometric/SB2 technique to an O-star binary. The 09.5\,Ib star
$\zeta$\,Ori Aa was found to have a mass of:\\[1mm] 
{\bf\boldmath \centerline
  {$M_{\rm \zeta\,Ori~Aa} = 24.8\pm5.6\,{\rm M}_\odot$} \unboldmath}\\[1mm]
We as well give a mass for its early B-type companion Ab, which is
16.4$_\odot$. This value is in reasonable agreement with the estimate of the
companion to be a little evolved very early B-type star. The mass of
$\zeta$\,Ori Aa is also in good agreement with the theoretically expected
value for its spectral type, i.e.~its track mass.

We stress that this is still a {preliminary determination}, based on only a
{few spectral lines} and not yet using all available spectroscopic data. Taking
advantage of the full wavelength range of the available {echelle data}, it
will be possible to determine the radial velocity curve of both components
with much {higher confidence}.
%


\end{document}